\documentclass[aps, twoside, nofootinbib]{revtex4}
\usepackage{amsfonts}
\usepackage{amssymb}
\usepackage[mathscr]{eucal}
\usepackage[dvips]{graphicx}

\def \be{\begin{equation}}
\def \ee{\end{equation}}
\def \bes{\begin{eqnarray}}
\def \ees{\end{eqnarray}}

\newcommand{\Cb}{{\rm \bf C}}

\def \sl2{SL(2,\Cb)}

\begin{document}
\title{A combinatorial and field theoretic path to quantum gravity: \\ the new challenges of group field theory}
\author{{\bf D. Oriti}\footnote{d.oriti@phys.uu.nl}}
\affiliation{\small Institute for Theoretical Physics and Spinoza
Institute \\ Utrecht University \\ Leuvenlaan 4, 3584 TD Utrecht,
The Netherlands}
\begin{abstract}
Group field theories are a new type of field theories over group
manifolds and a generalization of matrix models, that have
recently attracted much interest in quantum gravity research. They
represent a development of and a possible link between different
approaches such as loop quantum gravity and simplicial quantum
gravity. After a brief introduction to the GFT formalism we put
forward a long but still far from exhaustive list of open issues
that this line of research faces, and that could be represent
interesting challenges for mathematicians and mathematical
physicists alike.
\end{abstract}

\maketitle
\section{Introduction}

This article (as our talk at the workshop) has one main goal: to
draw attention on one recent approach to quantum gravity, named
group field theory, and on some of the many outstanding open
issues of this approach, in particular those that can be (in our
humble opinion) both mathematically attractive for those
mathematicians and mathematical physicists interested in
combinatorics and quantum field theory, as well as physically
crucial from a quantum gravity perspective. The hope and the
expectation (as stressed during the workshop) are therefore that
more mathematicians and mathematical physicists will join quantum
gravity theorists in the analysis and development of the group
field theory formalism, for mutual benefit and amusement, as well
as for speeding up progress towards a complete theory of quantum
gravity.

\medskip

Group field theories \cite{iogft, iogft2, laurentgft} can be
understood on the one hand as quantum field theories on particular
group manifolds, which are combinatorially non-local in a sense to
be clarified below, and on the other hand as a generalization of
matrix models of 2d quantum gravity. This generalization takes
place at two levels: in the added group structure and thus in the
number of dynamical degrees of freedom (GFTs are true quantum
field theories in contrast to the quantum mechanical matrix
models) and in the combinatorial structure of the Feynman diagrams
of the theory, arising in perturbative expansion (general
n-complexes as opposed to 2d ones). Indeed one can also see GFTs
as obtained by adding group structure to so-called tensor models
(see e.g. \cite{JanTensor}), in turn a purely combinatorial
generalization of matrix models. A much richer framework follows,
much of it being still unexplored from a mathematical as well as
physical perspective. The basic interpretative framework of group
field theory goes as follows: the GFT field, in models (aiming at)
describing D-dimensional quantum gravity, is interpreted as a
second quantized (D-1)-simplex, with (D-2)-faces of the same
labelled by group theoretic data, interpreted as (pre-)geometric
elementary quantities, or discrete quantum gravity variables.
Equivalently, the same data can be associated to the links of a
topologically dual graph, and the field is then seen as the second
quantization of a spin network functional \cite{LQG}. This means
that GFTs can be seen equivalently as a second quantized
formulation of spin network dynamics or as a field theory {\it of}
simplicial geometry.

\medskip

Already at this point one reason of interest in GFTs becomes
apparent: GFTs can potentially represent a common framework for
different current approaches to quantum gravity, in particular
canonical loop quantum gravity\cite{LQG} and simplicial quantum
gravity formalisms, namely quantum Regge calculus \cite{QRC} and
(causal) dynamical triangulations \cite{CDT}, because the same
mathematical structures that characterize these approaches also
enter necessarily and in very similar fashion in the GFT
framework. Let us be slightly more explicit here, even though all
of this will become clear later on when the GFT formalism will be
described in more detail. The connection with loop quantum gravity
arises first of all because GFT boundary states are given by (open
or closed) spin network states, i.e. graphs labelled by group
representations or group elements, which are indeed the
kinematical quantum states of gravity as discovered by loop
quantum gravity. Also, when the GFT partition function is expanded
in Feynman diagrams, they turn out to be given by spin foams, i.e.
labelled 2-complexes that first arose in the loop quantum gravity
context to represent the histories of spin network states, and the
Feynman amplitudes that are associated to them are nothing else
than spin foam models, i.e. combinatorial and algebraic sum over
histories first introduced to encode the dynamics of loop quantum
gravity states. The same Feynman diagrams, as we shall see,
identify simplicial complexes to which the GFT assigns geometric
data, weighted by amplitudes that can be derived from or related
to path integrals for simplicial gravity on the given complex.
Thus, for given Feynman diagram, the GFT provides a model
quantization of gravity in the spirit and language of quantum
Regge calculus, while, for given assignment of field degrees of
freedom (i.e. fixing the geometric data), the same GFT provides a
definition of the dynamics of quantum geometry via a sum over
triangulations (the perturbative expansion of the same GFT) in the
same spirit of the dynamical triangulations approach.

We refer the reader to the literature (especially \cite{iogft})
for a more extensive discussion on this. Here we stress only that,
this set of relations between different approaches to quantum
gravity, within the GFT formalism, suggests that any future
development and improved understanding of any aspect of GFTs is
likely to have implications and an impact in all of them. In this
sense, at the very least, GFTs may play a crucial role in current
quantum gravity research, in our opinion.

\section{The GFT formalism}
We now proceed to introduce the GFT formalism, in its main
features. Our treatment is going to be rather sketchy, and we
refer once more to the literature, in particular the reviews
\cite{iogft,iogft2,laurentgft}, for a more complete and detailed
treatment and a more extensive list of references.

\subsection{Kinematics}
We start from a field taken to be a
  $\mathbb{C}$-valued function of D group elements, for a generic
  group $G$, one for each of the D boundary (D-2)-faces of the
  (D-1)-simplex that the field $\phi$ represents: $$\phi(g_1,g_2,...,g_D):
  G^{\times D}\rightarrow \mathbb{C}.$$
We can identify the ordering of the arguments of the field with a
choice of orientation for the (D-1)-simplex it represents, and we
then encode the orientation properties of the corresponding
(D-1)-simplex in the complex structure by requiring invariance of
the field under even permutations $\sigma$ of its arguments (that
do not change the orientation) and trading odd permutations of
them with complex conjugation of the field. Given this symmetry,
then, a more precise definition of the field is:
$\phi(g_1,...,g_D)\equiv\sum_{\sigma}\phi(g_{\sigma(1)},...,g_{\sigma(D)})$.
Other symmetry properties can also be considered.

An additional symmetry that is usually imposed on the field (but
this is again model-dependent, of course) is the invariance under
diagonal action of the group $G$ on the D arguments of the field:
$\phi(g_1,...,g_D)=\phi(g_1g,...,g_Dg)$. This is the simplicial
counterpart of the Lorentz gauge invariance of continuum and
discrete first order gravity actions, and it has also the
geometric interpretation, at the simplicial level, of requiring
the D faces of a (D-1)-simplex to close to form the close
$S^{D-2}$ surface representing its boundary. This reduces the
number of degrees of freedom from those represented by the group
elements associated to the $(D-2)$ faces, but at the same time
intertwines them in a rather non-trivial way, as it is clear by
the resulting more complicated geometry and topology of
configuration space.

As in any other field theory, a momentum representation for the
field and its dynamics is also available and is obtained by
harmonic analysis on the group manifold $G$. The field can be
expanded in modes as:
$$\phi(g_i)=\sum_{J_i,\Lambda,k_i}\phi^{J_i\Lambda}_{k_i}\left( \prod_iD^{J_i}_{k_il_i}(g_i)\right) C^{J_1..J_D\Lambda}_{l_1..l_D}, $$ with the $J$'s
labelling representations of $G$, the $k$'s vector indices in the
representation spaces, and the $C$'s being intertwiners of the
group $G$. We have labelled an orthonormal basis of intertwiners
by an extra parameter $\Lambda$ (depending on the group chosen and
on the dimension D, this may actually be a shorthand notation for
{\it a set} of parameters). That this decomposition is possible is
not guaranteed in general (as the harmonic analysis of non-compact
groups, for example, may not be under control), but it is in fact
true for all the known quantum gravity GFT models, which are based
on the Lorentz group or on extensions of it. A geometric
interpretation of the field variables is obtained either looking
at the Feynman amplitudes for the GFT at hand, in turn usually
obtained from a discretization of some continuum gravity action,
or from the direct (1st) quantization of simplicial structures,
e.g. by geometric quantization methods. The group variables are
then seen to represent parallel transport of a (gravity)
connection along elementary paths dual to the (D-2)-faces, and the
representations $J$ are usually put in correspondence with the
volumes of the same (D-2)-faces, the details of this
correspondence depending, however, on the specific model.
\begin{figure}[t]
\includegraphics[width=15.4cm, height=3cm]{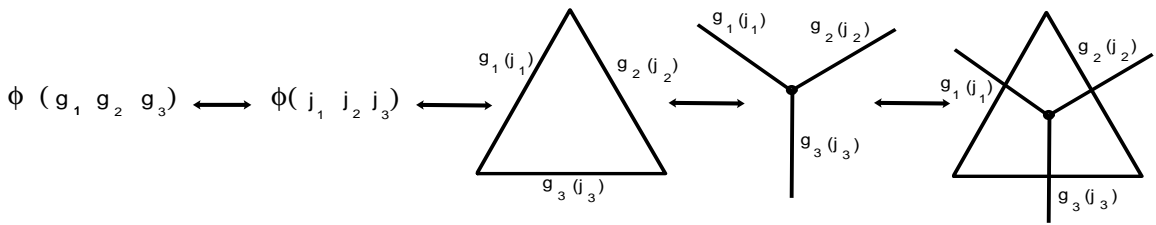}
\caption{For the $D=3$ case, the association of a field with a
2-simplex, a triangle, or equivalently its dual vertex, and of its
arguments with the edges (1-faces) of the triangle, or
equivalently with the links incident to the vertex, together with
the consequent labelling of the by group-theoretic variables.}
\end{figure}

Just as one identifies a single field (or the corresponding 1st
quantized wave function) with a single (D-1)-simplex, a simplicial
space built out of $N$ such (D-1)-simplices is described by the
tensor product of $N$ such wave functions, with suitable
constraints implementing the fact that some of their (D-2)-faces
are identified. For example, a state describing two
(D-1)-simplices glued along one common (D-2)-face would be
represented by: $\phi^{J_1 J_2..J_D\Lambda}_{k_1 k_2...k_D}
\phi^{\tilde{J}_1
  J_2...\tilde{J}_D\tilde{\Lambda}}_{\tilde{k}_1 k_2...\tilde{k}_D}$,
where the gluing is along the face labelled by the representation
$J_2$, and effected by the contraction of the corresponding vector
indices (of course, states corresponding to disjoint
(D-1)-simplices are also allowed).

\begin{figure}
\includegraphics[width=12.8cm, height=3cm]{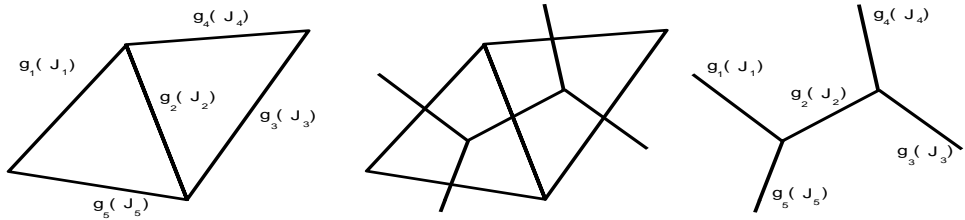}

\caption{A \lq 2-particle state\rq (again, in the D=3 example)}
\end{figure}

We see that states of the theory are then labelled, in momentum
space, by {\it spin networks} of the group $G$\cite{LQG}. The
corresponding second quantized theory is a GFT. Spin networks are
also relevant for the construction of GFT observables. These are
given \cite{laurentgft} by gauge invariant functionals of the GFT
field, and can be constructed in momentum space using spin
networks according to the formula:
$$ O_{\Psi=(\gamma, j_e,i_v)}(\phi)=\left(\prod_{(ij)}\int dg_{ij}dg_{ji}\right) \Psi_{(\gamma, j_e,i_v)}(g_{ij}g_{ji}^{-1})\prod_i \phi(g_{ij}),$$
where  $\Psi_{(\gamma, j_e,i_v)}(g)$ identifies a spin network
functional \cite{LQG} for the spin network labelled by a graph
$\gamma$ with representations $j_e$ associated to its edges and
intertwiners $i_v$ associated to its vertices, and $g_{ij}$ are
group elements associated to the edges $(ij)$ of $\gamma$ that
meet at the vertex $i$.

\subsection{Dynamics}
On the basis of the above kinematical structure, one aims at
defining a field theory for describing the interaction of
fundamental building blocks of space ((D-1)-simplices or spin
network vertices), and in which a typical interaction process will
be characterized by a D-dimensional simplicial complex. In the
dual picture, the same will be represented as a spin foam (
labelled 2-complex). This is the straightforward generalization of
the way in which 2d discretized surfaces emerge from the
interaction of matrices (graphically, segments), or ordinary
Feynman graphs emerge from the interaction of point particles. A
{\it discrete} spacetime emerge then from the theory as a virtual
construct, a possible interaction process among the quanta of the
theory.

In order for this to be realized, the classical field action in
group field theories has to be chosen appropriately. In this
choice lies the main peculiarity of GFTs with respect to ordinary
field theories. This action, in configuration space, has the
general QFT structure:

\begin{eqnarray} S_D(\phi, \lambda)= \frac{1}{2}\left(\prod_{i=1}^D\int
  dg_id\tilde{g}_i\right)
  \phi(g_i)\mathcal{K}(g_i\tilde{g}_i^{-1})\phi(\tilde{g}_i)
  +
  \frac{\lambda}{(D+1)!}\left(\prod_{i\neq j =1}^{D+1}\int dg_{ij}\right)
  \phi(g_{1j})...\phi(g_{D+1 j})\,\mathcal{V}(g_{ij}g_{ji}^{-1}), \label{eq:action}
\end{eqnarray}
and it is of course the choice of kinetic and interaction
functions $\mathcal{K}$ and $\mathcal{V}$ that define the specific
model considered. Obviously, the same action can be written in
momentum space after harmonic decomposition on the group manifold.

The mentioned peculiarity is in the combinatorial structure of the
pairing of field arguments in the kinetic and vertex terms, as
well as from their degree as polynomials in the field.  The
interaction term describes the interaction of D+1 (D-1)-simplices
(the fundamental \lq atoms of space\rq) to form a D-simplex (\lq
the fundamental virtual \lq atom of spacetime\rq) by gluing along
their (D-2)-faces (arguments of the fields), that are {\it
pairwise} linked by the interaction vertex. The nature of this
interaction is specified by the choice of function $\mathcal{V}$.
The (quadratic) kinetic term involves two fields each representing
a given (D-1)-simplex seen from one of the two D-simplices
(interaction vertices) sharing it, so that the choice of kinetic
functions $\mathcal{K}$ specifies how the information and
therefore the geometric degrees of freedom corresponding to their
D (D-2)-faces are propagated from one vertex of interaction
(fundamental spacetime event) to another. Let us mention here that
one can consider generalizations of the above combinatorial
structure, for example defining vertex functions whose
combinatorics corresponds to the gluing of (D-1)-simplices to form
different sorts of D-dimensional complexes (e.g. hypercubes etc).
We are not going to elaborate further on this.

Because of the peculiar way in which field arguments are paired in
the interaction term, we may consider GFTs as {\it combinatorially
non-local field theories} (as opposed to theories in which the
nonlocality is originated by higher derivatives terms, for
example). Before detailing more the interesting structure of
Feynman diagrams resulting from this combinatorial nonlocality,
let us stress once more that it is basically in it that lies the
peculiarity of GFTs as field theories. Indeed, as for the rest, we
have an almost ordinary field theory, in that we can rely on a
fixed background metric structure, given by the invariant
Killing-Cartan metric on the group manifold, a fixed topology,
given again by the topology of the group manifold, the usual
splitting between kinetic (quadratic) and interaction (higher
order) term in the action, that will later allow for a
straightforward perturbative expansion, and the usual conjugate
pictures of configuration and momentum space. This allows us to
use all usual QFT techniques and language in the analysis of GFTs,
and thus of quantum gravity, even though we remain in a background
independent (in the physical sense of \lq spacetime
independent\rq) context. The importance of this, in a
non-perturbative quantum gravity framework, should not be
underestimated, we think.

\ \

Let us now turn to the quantum dynamics. Most of the research in
this area has concerned the perturbative aspects of this dynamics
around the no-particle state, the complete vacuum, and the main
guide for model building have been, up to now, only the properties
of the resulting Feynman amplitudes. This relevant Feynman
expansion is:

$$ Z\,=\,\int
\mathcal{D}\phi\,e^{-S[\phi]}\,=\,\sum_{\Gamma}\,\frac{\lambda^{N_v(\Gamma)}}{sym[\Gamma]}\,Z(\Gamma),
$$
where $N_v$ is the number of interaction vertices $v$ in the
Feynman diagram $\Gamma$, $sym[\Gamma]$ is the number of
automorphisms of $\Gamma$ and $Z(\Gamma)$ the corresponding
Feynman amplitude. Each edge of the Feynman graph is made of $D$
strands, one for each argument of the field\footnote{One could
write explicitly down and keep track of each of the spacetime
coordinates on which the field depends also in ordinary QFT in
Minkowski space, but it would be a pedantic and rather useless
complication. The need to keep track of each argument of the
field, in a GFT context, comes from the mentioned nonlocality of
the interaction term, which is at the origin of the
combinatorially nontrivial structure of the resulting Feynman
diagrams.} and each one is then re-routed at the interaction
vertex, with the combinatorial structure of an $D$-simplex,
following the pairing of field arguments in the vertex operator.
Diagrammatically:

\begin{figure}[here]
\setlength{\unitlength}{1cm}
\begin{minipage}[t]{3.5cm}
\hspace{-0.3cm}\includegraphics[width=3.5cm,
height=2.5cm]{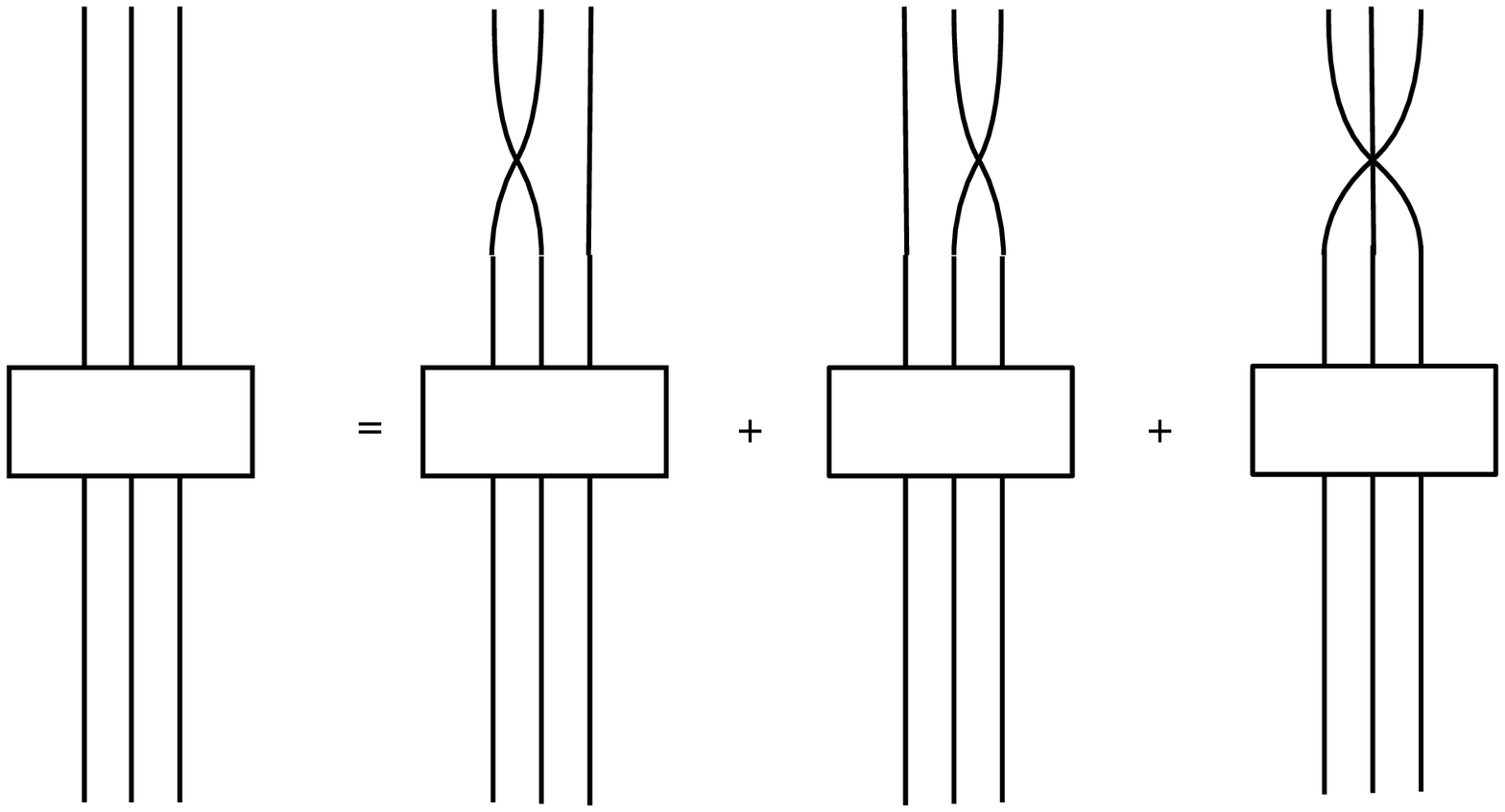}
\end{minipage}
\hspace{0.3cm}
\begin{minipage}[t]{5.5cm}
\includegraphics[width=3.5cm, height=2.5cm]{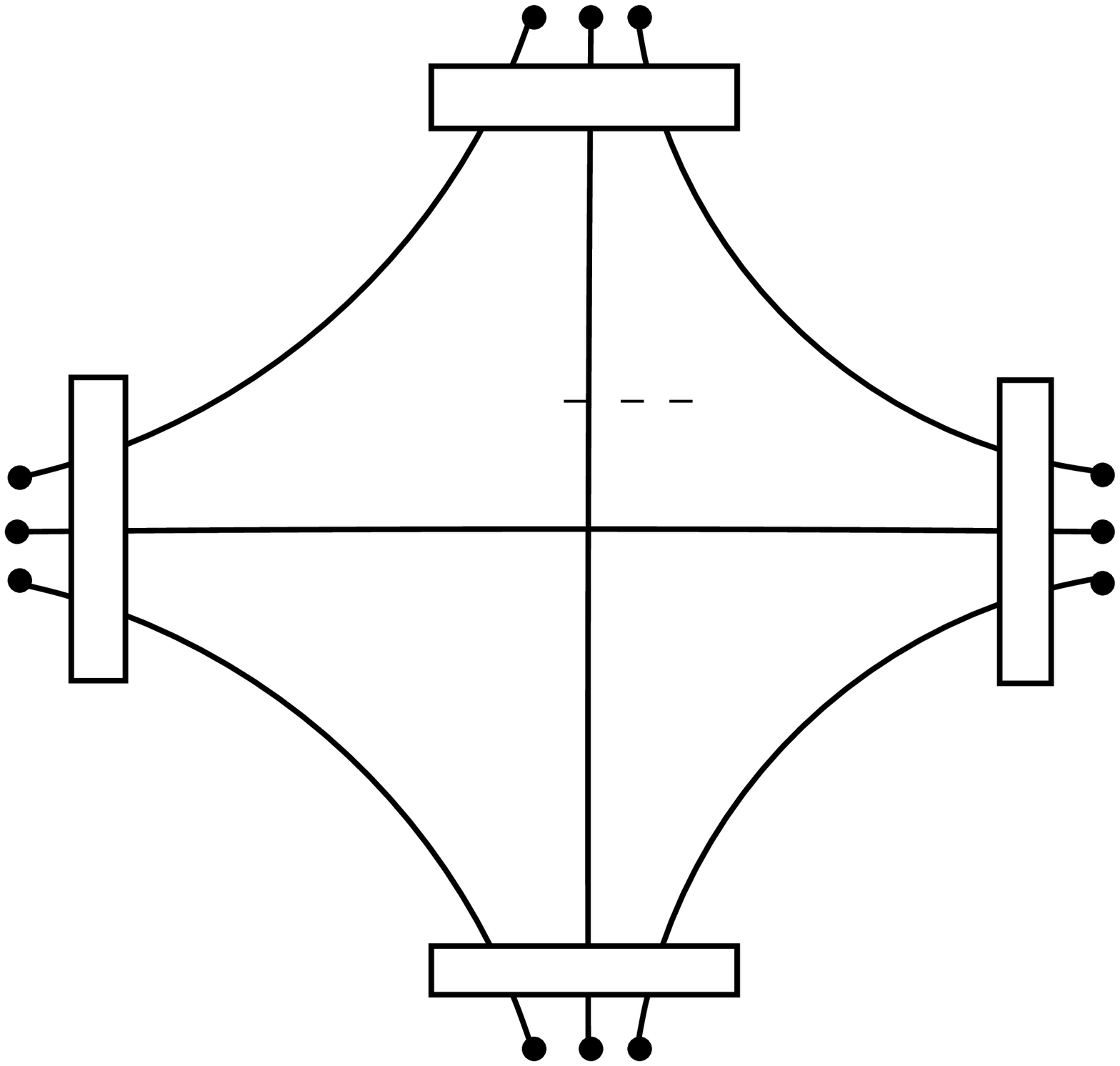}
\end{minipage}
\hspace{0.1cm}
\begin{minipage}[t]{5cm}
\includegraphics[width=7cm, height=3cm]{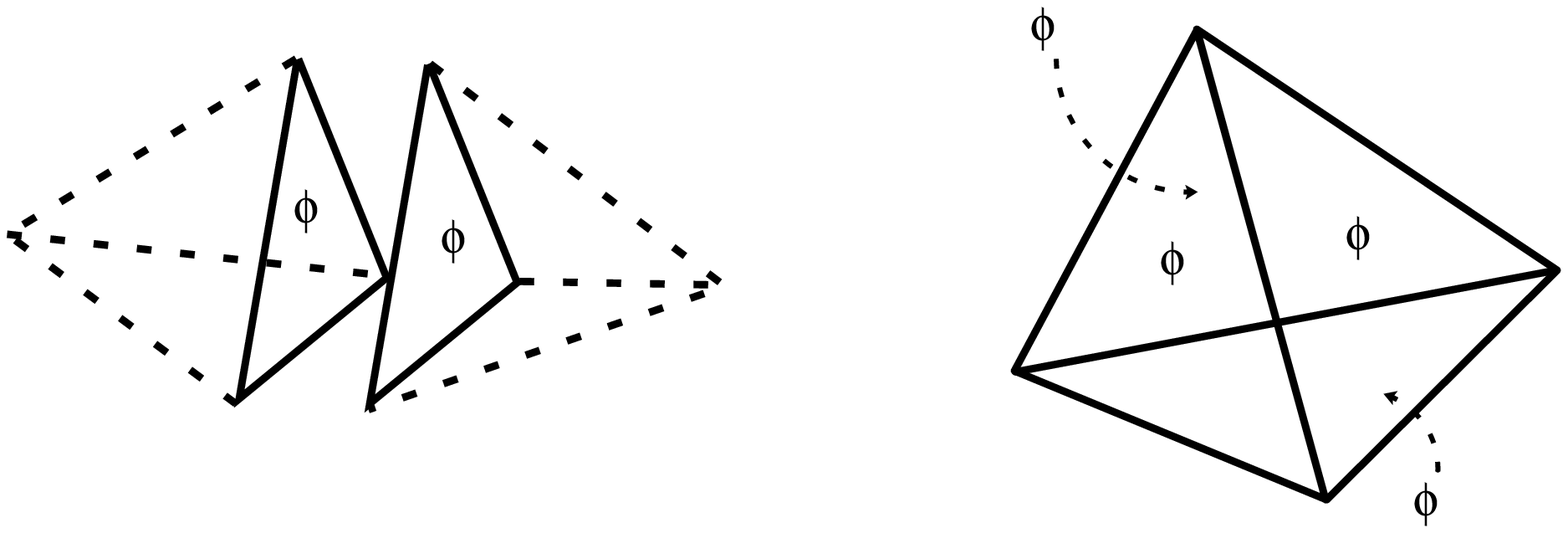}
\end{minipage}

\end{figure}

Each strand in an edge of the Feynman diagram goes through several
vertices, coming back where it started, for closed Feynman
diagrams, and therefore identifies a 2-cell (for open graphs, it
may end up on the boundary, identifying then an open 2-cell). Each
Feynman diagram $\Gamma$ is then a collection of 2-cells (faces),
edges and vertices, i.e. a 2-complex, that, because of the chosen
combinatorics for the arguments of the field in the action, is
topologically dual to a D-dimensional simplicial complex. Notice
that the resulting 2-cells can be glued (i.e. can share edges) in
all sorts of ways, forming for example \lq\lq bubbles\rq\rq, i.e.
closed 3-cells.

No restriction on the topology of the resulting diagram/complex is
imposed, a priori, in their construction, so the resulting
complexes/triangulations can have arbitrary topology. Each
resulting 2-complex or triangulation corresponds to a particular
{\it scattering
  process} of the fundamental building blocks of space,
i.e. (D-1)-simplices. Each line of propagation, made as we said
out of D strands, is labelled, on top of the group/representation
data, by a permutation of $(1,..,D)$, representing the labelling
of the field variables, and all these data are summed over in the
construction of the Feynman expansion. The sum over permutations
affects directly the combinatorics of the allowed gluings of
vertices with propagators. The above choice of permutation
symmetry for the field (with the orientation encoded in the
complex structure) implies that only {\it even} permutation appear
as labellings of propagation lines. In turn, this ensures that
only orientable complexes are generated in the Feynman expansion
of the field theory (see \cite{DP-P} for a more detailed
treatment). The more restrictive choice of invariance of the field
under {\it any} permutation of its arguments results, for example,
in the presence of non-orientable complexes as well in the Feynman
expansion.
\begin{figure}[here]
\includegraphics[width=12.5cm, height=3cm]{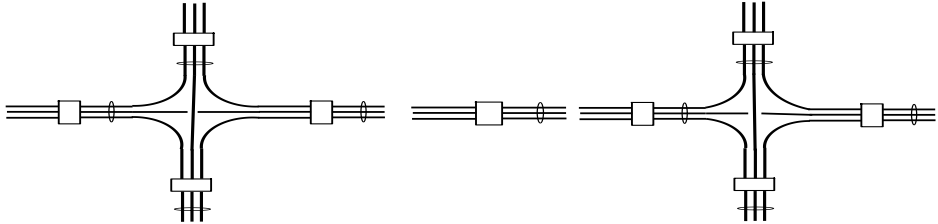}
\caption{The gluing of vertices of interaction through
propagators, again in the D=3 example. The rectangles represent
the additional integrations imposing gauge invariance under the
action of $G$, while the ellipses represent the implicit sum over
permutations of the (labels of the) strands to be glued.}
\end{figure}

\begin{figure}[here]
\includegraphics[width=12cm, height=6cm]{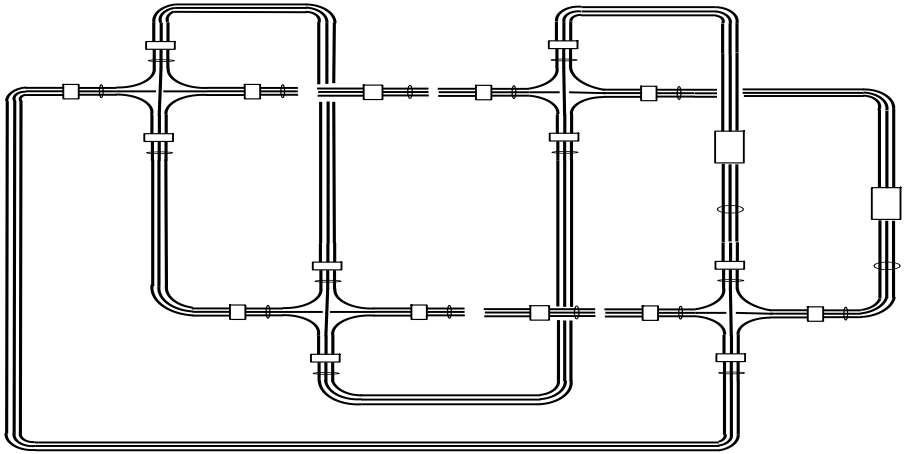}
\caption{An example, in D=3, of a closed GFT Feynman diagram, with
4 vertices and 8 propagators.}
\end{figure}

As said, each strand in a propagation line carries a field
variable, i.e. a group element in configuration space or a
representation label in momentum space. After the closure of the
strand to form a 2-cell in a closed diagram, the same
representation label ends up being associated to this 2-cell.
Therefore in momentum space each Feynman graph is given by a spin
foam (a 2-complex with faces labelled by representation
variables), and each Feynman amplitude (a complex function of the
representation labels, obtained by contracting vertex amplitudes
with propagator functions) by a so-called spin foam model
\cite{SF}:

$$
Z(\Gamma)=\sum_{J_f}\prod_f A_f(J_f) \prod_e A_e(J_{f\mid e})
\prod_v A_v(J_{f\mid v}),
$$
where we have highlighted the fact that the amplitudes can be
factori  Given the mentioned geometric interpretation of the
representation variables (edge lengths, areas, etc) (see
\cite{SF,iogft}), each of these Feynman amplitudes corresponds to
a definition of a sum-over-histories for discrete quantum gravity
on the specific triangulation dual to the Feynman graph, although
the quantum amplitudes for each geometric configuration are not
necessarily given by the exponential of a discrete gravity action.

One can show that the inverse is also true: any local spin foam
model can be obtained from a GFT perturbative expansion
\cite{carlomike,laurentgft} (even though this does not imply that
the reconstruction of the underlying GFT action from the knowledge
of the spin foam amplitudes is immediate nor easy). This implies
on the one hand that the GFT approach incorporates the spin foam
in its perturbative aspects, and on the other hand that it goes
potentially far beyond it, since there is of course much more in a
QFT than its perturbative expansion. The sum over Feynman graphs
gives then a sum over spin foams, and equivalently a sum over
triangulations, augmented by a sum over algebraic data (group
elements or representations) with a geometric interpretation,
assigned to each triangulation. This perturbative expansion of the
partition function also allows for a perturbative evaluation of
expectation values of GFT observables, as in ordinary QFT. In
particular, the transition amplitude (probability amplitude for a
certain scattering process) between certain boundary data
represented by two spin networks, of arbitrary combinatorial
complexity, can be expressed as the expectation value of the field
operators having the same combinatorial structure of the two spin
networks \cite{laurentgft, iogft}.

$$
\langle \Psi_1\mid\Psi_2\rangle = \int
\mathcal{D}\phi\,O_{\Psi_1}\,O_{\Psi_2}\,e^{-S(\phi)} =
\sum_{\Gamma/\partial\Gamma=\gamma_{\Psi_1}\cup\gamma_{\Psi_2}}\,\frac{\lambda^N}{sym[\Gamma]}\,Z(\Gamma)
$$
where the sum involves only 2-complexes (spin foams) with boundary
given by the two spin networks chosen.

\ \

The above perturbative expansion involves thus two types of sums:
one is the sum over geometric data (group elements or
representations of $G$) entering the definition of the Feynman
amplitudes as the GFT analogue of the integral over momenta or
positions of usual QFT; the other is the overall sum over Feynman
diagrams. We stress again that, in absence of additional
restrictions being imposed on the GFT, the last sum includes a sum
over all triangulations for a given topology and a sum over all
topologies\footnote{There is no algorithmic procedure that allows
to distinguish topologies in $D\geq 3$, so we cannot partition
this sum into the two sub-sums mentioned. What ensures us that all
topologies are present in the full sum is, however, that any
simplicial complex is obtained by an appropriate gluing and face
identification of fundamental simplicial building blocks, and that
all possible such gluings and identifications are included in the
GFT Feynman expansion.}.

\subsection{Examples}
We now give a few examples of specific GFT models, again referring
to the literature for more details.

\subsubsection{D=2 and matrix models}
The easiest example is a straightforward generalization of matrix
models for 2d quantum gravity to a GFT, obtained by adding group
structure to them, but keeping the same combinatorics, and it is
given:
\begin{equation}
S[\phi]=\int_G dg_1dg_2 \frac{1}{2}\phi(g_1,g_2)\phi(g_1,g_2) +
\frac{\lambda}{3!}\int
dg_1dg_2dg_3\phi(g_1,g_2)\phi(g_1,g_3)\phi(g_2,g_3)
\end{equation}
where $G$ is a generic compact group, say $SU(2)$, and the
symmetries mentioned above are imposed on the field $\phi$
implying, in this case: $\phi(g_1,g_2)=\tilde{\phi}(g_1g_2^{-1})$.
The relation with matrix models is apparent in momentum space,
expanding the field in representations $j$ of $G$ to give:

\begin{equation}
S[\tilde{\phi}]=\sum_j \textit{dim}(j) \left(
\frac{1}{2}tr{(\tilde{\phi}_j^2)}
+\frac{\lambda}{3!}tr(\tilde{\phi}_j^3)\right) \label{mm}
\end{equation}

where the field modes $\tilde{\phi}_j$ are indeed matrices with
dimension $\textit{dim}(j)$. Thus, one gets a sum of matrix models
actions of increasing dimensions. Alternatively, one can see the
above as the action for a single matrix model in which the
dimension of the matrices has been turned from a parameter into a
dynamical variable.

\subsubsection{D=3 and topological BF theory}
Another much studied example (actually the best understood one) is
the group field theory, first proposed by D. Boulatov, for
topological BF theory in 3d, which in turn is closely related to
3-dimensional quantum gravity in first order formalism.

Kinetic and vertex terms are then chosen as follows:

\be \mathcal{K}(g_i,\tilde{g}_i) = \int_G dg \prod_i
\delta(g_i\tilde{g}_i^{-1}g),\;\;\;\;\;\;\mathcal{V}(g_{ij},g_{ji})
= \prod_i\int_G dg_i \prod_{i<j}\delta(g_i
g_{ij}g_{ji}^{-1}g_j^{-1}), \label{PR} \ee

where the integrals impose the gauge invariance under the action
of $G$. The choice of $G=SO(3)$ or $G=SO(2,1)$ provides a
quantization of 3d gravity in the Euclidean and Minkowskian
signatures, respectively, and the Feynman amplitudes are given by
the so-called Ponzano-Regge spin foam model. The choice of the
quantum group $SU(2)_q$ gives the Turaev-Viro topological
invariant, conjectured to correspond, when $q$ is a root of unity,
to 3d gravity with positive cosmological constant. For the $SU(2)$
case, and for real field, the action is then:
\begin{eqnarray}
S[\phi]=\prod_i \int_{SU(2)} \phi(g_1,g_2,g_3)\phi(g_1,g_2,g_3)+
\frac{\lambda}{4!}\prod_{i=1}^{6}\int_{SU(2)}dg_i
\;\phi(g_1,g_2,g_3)\phi(g_3,g_4,g_5)\phi(g_5,g_2,g_6)\phi(g_6,g_4,g_1),
\;\;\;\;\end{eqnarray} and the Feynman amplitudes are:

$$
Z(\Gamma)\,=\, \left(\prod_{e*\in \Gamma} \int d g_{e*}\right)
\,\prod_{f*}\,\delta (\prod_{e*\in\partial f*} g_{e*} )
$$
where $e*$ are dual edges of the 2-complex and $f*$ its 2-cells.
This is the same quantity one obtains from a path integral
quantization of discretized BF theory on the triangulation dual to
$\Gamma$, confirming the above-given interpretation.

Lots is known about the last model, including the appropriate
gauge fixing procedure of its Feynman amplitudes, the coupling of
matter fields, etc. and we refer once more to the literature for
more details, starting from \cite{PR1}.

\subsubsection{Models with additional structure}
One may then consider models with additional structure and more
involved kinetic and vertex operators. One example are the
generalized models of \cite{generalised}, based, in the D=3 case,
on a complex field on $(G\times \mathbb{R})^{\times 3}$, but with
a $G$-only diagonal invariance, with $G=SU(2)$ or
$G=SL(2,\mathbb{R})$ in the Riemannian and Lorentzian cases
respectively. The kinetic and vertex operators are then:

\begin{eqnarray} \hspace{-0.5cm} \mathcal{K}(g_i,s_i, \tilde{g}_i,\tilde{s}_i) = \sum_{\sigma}\int dg  \,
\prod_i\,\left( i\partial_{s_{i}} + \square_i \right)\delta(g_i g
\tilde{g}_{\sigma(i)}^{-1})\delta(s_i -
\tilde{s}_{\sigma(i)}^{-1}) \nonumber \\ \hspace{-2cm}
\mathcal{V}(g_{ij},s_{ij}) =\,\prod_{i\neq j}^{} \delta(g_{ij}
g_{ji}^{-1})\delta(s_{ij} - s_{ji}) \label{EFnr}
\end{eqnarray}
where $g_i\in G$, $s_i\in\mathbb{R}$ and $\square$ is the
Laplace-Beltrami operator on $G$.

One more example of a similar type, a sort of \lq relativistic\rq
upgrading of the previous one, and that is currently being studied
and developed \cite{iotim}, as the possible explicit common ground
for loop quantum gravity and simplicial quantum gravity, uses
instead a complex field over $(G\times \mathbb{R}^3)^{\times 3}$
and the kinetic and vertex terms:

\begin{eqnarray}
\hspace{-0.5cm} \mathcal{K}(g_i,x_i, \tilde{g}_i,\tilde{x}_i) =
\sum_{\sigma}\int dg  \, \prod_i\,\left( \triangle_{i} + \square_i
\right)\delta(g_i g \tilde{g}_{\sigma(i)}^{-1})\delta(x_i -
\tilde{x}_{\sigma(i)}^{-1}) \nonumber
\\ \hspace{-2cm} \mathcal{V}(g_{ij},x_{ij}) =\,\prod_{i\neq j}^{}
\delta(g_{ij} g_{ji}^{-1})\delta(x_{ij} - x_{ji}) \label{EF}
\end{eqnarray}
where $g_i\in G$, $x_i\in\mathbb{R}^3$, $\triangle$ is now the
Laplace-Beltrami on $\mathbb{R}^3$ and $\square$ is again the
Laplace-Beltrami on $G$. The interest in this last model lies in
the fact that the corresponding Feynman amplitudes have {\it
exactly} the form of path integrals for simplicial quantum
gravity, i.e. quantum Regge calculus, in 1st order form . Similar
models exist in higher dimension ($D\geq 3$)\cite{iotim}.

\section{A (far from comprehensive) list of open issues}
We now outline, in a rather schematic way, a list of open issues
within the group field theory approach, that are in our opinion of
considerable mathematical interest as well as crucial for a better
physical understanding of this class of models. We stress once
more that this list is, by all means, far from exhaustive, even
from a purely mathematical perspective. At the same time, we also
stress that it is only for the stated aims of this article that we
do not review or discuss more what, instead, {\it is} known about
GFTs, and the many results already obtained in this approach. For
an account of them, we refer to the cited literature \cite{iogft,
iogft2, laurentgft}.

\subsection{Classical solutions and their relevance}
GFTs were historically developed from the perspective of either
loop quantum gravity and spin foam models or dynamical
triangulations, and thus the focus of attention has always been
their perturbative structure and the properties of their Feynman
amplitudes. However, changing slightly the perspective on them,
and seeing them just as field theories (of a peculiar type), the
first thing one would think of analyzing is their classical
structure. In particular, one would start from their classical
equations of motion. The importance of these equations from the
GFT point of view are obvious: they define the classical dynamics
of the field theory, they identify of classical background
configurations around which to expand in a semi-classical
perturbative definition of the path integral, etc. Their relevance
from a quantum gravity perspective is easily understood if one
recalls the interpretation of GFTs as second quantized theories of
simplicial geometry or of spin network states. Then the GFT
classical action should encode the full 1st quantized dynamics,
and the classical equations of motion should correspond to the
full 1st quantized wave function equations. In the specific case
of gravity, then, solving the GFT classical equations means
identifying non-trivial quantum gravity wave functions satisfying
{\bf all} the quantum gravity constraints, an important and still
unachieved goal of canonical quantum gravity, even in the modern
loop quantized formalism.

\medskip

The classical equations of motion following from the general form
of the action \ref{eq:action} (considering for simplicity a real
field) are:
$$
\prod_j\int
d\tilde{g}_{1j}\mathcal{K}(g_{1j}\tilde{g}_{1j}^{-1})\phi(\tilde{g}_{1j})
+
  \frac{\lambda}{D!}\left(\prod_{j=1}^{D+1}\int dg_{2j}..dg_{D+1 j}\right)
  \phi(g_{2j})...\phi(g_{D+1 j})\,\mathcal{V}(\{g_{ij}\})=0 .
$$

For a generic differential operator $\mathcal{K}$, they are then
rather complicated non-linear equations of integro-differential
type, with the main complications coming once more from the
particular pairing of variables in the interaction term. No
detailed analysis of these equations in any specific GFT model nor
of their solutions has been carried out to date.

\medskip

Solving these equations in complete generality, obtaining a
complete classification of their exact solutions, is probably
beyond reach, even in low D. However, two more modest goals are
probably within reach, and both would be of importance.

1) Identify at least {\it some} exact solutions of the full GFT
equations, for some currently studied GFT model, and analyze their
mathematical structure and physical meaning. Most likely, this has
to be done first in symmetry reduced cases, and, on top of
identifying some particularly simple solution of a given GFT
model, it would be of considerable interest to develop a general
theory of symmetry reduction for GFTs, to be then applied to the
various models.

2) Develop an exact formalism for obtaining {\it approximate}
solutions to the above field equations, e.g. in perturbation
expansion in the coupling parameter $\lambda$, or in some other
parameter that could allow to take into account both the
non-linearity of the equations and the non-local nature of the
field coupling in a perturbative way.

Indeed, the classical solutions of the (local and linear) free
field theory equations
$$ \left( \prod_j\int
d\tilde{g}_{j}\right)\mathcal{K}(g_{j}\tilde{g}_{j}^{-1})\phi(\tilde{g}_{j})=0$$
are rather easily obtained, but at the same time are not very
interesting physically, for most known models.

\medskip

Let us now given one example of a GFT equation of motion, to
clarify some of the above issues. Take the, arguably simplest, GFT
model \ref{PR}; the corresponding equations are:

$$
\int d\tilde{g}\,\phi(g_1\tilde{g},g_2\tilde{g},g_3\tilde{g}) +
\,\lambda\,\prod_{i=1}^{3} \int d\tilde{g}_{i}\prod_{j=4}^{6}\int
dg_j\, \phi(g_3\tilde{g}_1,g_4\tilde{g}_1,g_5\tilde{g}_1)
\times\phi(g_5\tilde{g}_2,g_6\tilde{g}_2,g_2\tilde{g}_2)\phi(g_6\tilde{g}_3,
g_4\tilde{g}_3, g_1\tilde{g}_3)\,=\,0
$$
where we have included additional integrals implementing the gauge
invariance. We do not know the general solution of this equation,
nor any exact method for obtaining it or any exact procedure for
approximating it. We do know some exact simplified (families of)
solutions \cite{winstonetera}, though, which are the following:

$$
\phi_f(g_1,g_2,g_3)=\sqrt{\frac{3!}{-\lambda}}\,\int dg \,
\delta(g_1 g)\, f(g_2 g)\, \delta(g_3 g),
$$
parametrized by a real function $f$ on $SU(2)$. More solutions of
the same type can be constructed, even in higher dimensions.

However, there is a lot yet to be clarified concerning the
physical meaning of these simple known solutions, and how their
mathematical structure implements it. We refer in particular to
the issue of topology change. Let us clarify. As discussed above,
the GFT \ref{PR} provides a quantization of topological BF theory
in 3d, which takes into account topology changing configurations,
as it is clear in perturbative expansion. Accordingly, {\it any}
classical solution should be interpreted as the wave function for
a {\it flat} geometry and some prescribed dynamics for the {\it
topological} degrees of freedom. How a flat geometry is
implemented in the above solutions, and which prescription for the
dynamics for space topology is encoded in them, as well as how the
two ingredients are intertwined, is mathematically yet to be
understood. Same applies to other known simple solutions.

The same issues, from the purely mathematical ones to those
concerning the physical interpretation, are of course even more
complex and interesting at the same time in the $D=4$ case, for
non-topological models, where geometry itself has a highly
non-trivial dynamics. For example, the model \ref{EF}, which
presents already a set of non-trivial free field solutions (e.g.
(the group analogue of) plane waves on both $\mathbb{R}^3$ and
$G$).

\subsection{Classical and quantum Hamiltonian analysis}
Still sticking to a purely formal field theoretic perspective on
GFTs, i.e. keeping aside momentarily the quantum gravity
interpretation of the same, one more basic aspect of them is both
undeveloped, and mathematically challenging: their
canonical/Hamiltonian formulation. A general formalism for the
canonical analysis of GFTs at both classical and quantum levels
has been recently proposed \cite{ioJimmy}, and applied to the
model \ref{EFnr}, but it represents, especially at the
mathematical level, only a first step.

What are the main difficulties involved in this analysis for GFTs?

Once more, they stem from the peculiar combinatorial structure of
the GFT action, but also from the need, resulting from the quantum
gravity interpretation, to deal with all of the field arguments on
equal footing, since they propagate almost independently from each
other (due again to the combinatorics in the vertex).

Let us be more explicit. Consider the action with kinetic and
vertex term \ref{EFnr}, and restrict the attention to the kinetic
term only (determining the symplectic structure of the theory and
thus providing the basis for the hamiltonian analysis):

$$ S=\left( \prod_i \int_G dg_i
\int_\mathbb{R} ds_i\right) \,
\phi^\dagger(g_1,s_1;...;g_D,s_D)\prod_i\,\left( i\partial_{s_{i}}
+ \square_i \right)\phi(g_1,s_1;...;g_D,s_D) + h.c.$$ for generic
group $G$ (Riemannian or Lorentzian).

The kinetic term has the structure of a product of differential
operators, each acting independently on one of the D (sets of)
arguments of the field. Each of them is a Schroedinger-like
operator with \lq \lq Hamiltonian\rq\rq $\square$. This suggests
that one should consider the variables $s_i$ as \lq\lq time\rq\rq
variables, to be used in a GFT generalization of the usual
time+space splitting of the configuration space coordinates, with
the group elements treated instead as \lq\lq space\rq\rq. This is
reasonable (and indeed the correct thing to do), but it implies we
have a field theory with D \lq\lq times\rq\rq, all to be treated
on equal footing. In turn this immediately implies that the naive
phase space has coordinates $(\phi, \pi_i = \frac{\delta
L}{\delta\partial_{s_i}\phi})$, with one field and D conjugate
momenta. Clearly, a generalization of usual Hamiltonian mechanics
is needed. One approach, and the one chosen in \cite{ioJimmy}, is
to use the DeDonder-Weyl generalized Hamiltonian mechanics, as
developed at both the classical and quantum level by Kanatchikov
\cite{kanatchikov}, as a starting point and to adapt it to the
peculiar GFT setting. The framework chosen is thus that of {\it
polysymplectic (or polymomentum) mechanics} \cite{kanatchikov},
and we refer to the literature for more mathematical details on
this beautiful formalism.

The general idea of how to adapt this general formalism to the GFT
case is the following. One starts from a \lq\lq covariant\rq\rq
definition of momenta, hamiltonian density, Poisson brackets, etc
treating all \lq\lq time variables\rq\rq on equal footing at
first, i.e. when defining densities. Then one defines the 'scalar'
quantities referring to each \lq time direction\rq (to be turned
into operators at the quantum level), including a set of D
Hamiltonians, by integration over appropriate hypersurfaces in
$(G\times\mathbb{R})^{\times D}$, so that each Hamiltonian refers
to a single time direction, but at the same time all time
directions are treated equally but independently, corresponding to
an equal but independent evolution (propagation) of the
corresponding degrees of freedom. A similar procedure is adopted
for other canonical quantities, e.g. Poisson brackets, scalar
products etc.

Let us sketch one example of such procedure, for the case $D=2$
and the above choice of GFT action, outlining only the definition
of the Hamiltonians, in order to convey at least the flavour of
the formalism. We refer to \cite{ioJimmy} for the full treatment
of this case as well as for a full, but still in many ways
preliminary, description of the GFT Hamiltonian analysis, while we
refer to \cite{kanatchikov} for the complete exposition of the
state of the art in polymomentum Hamiltonian mechanics.

We start from the naive phase space
$(\phi,\phi^\dagger,\pi_\phi^i=\frac{\delta
L}{\delta\partial_{s_i}\phi},\pi_{\phi^\dagger}^i=\frac{\delta
L}{\delta\partial_{s_i}\phi^\dagger})$ (we call it \lq naive\rq
simply because in the polymomentum formalism phase space variables
are actually differential forms on spacetime, while the $\pi$
above are not), with the product structure of the kinetic term
resulting in a peculiar expression for the momenta, e.g.
$\pi_\phi^1 = (-i\partial_2 +\square_2)\phi^\dagger$, and define
the DeDonder-Weyl Hamiltonian density (summation over repeated
indices understood):

$$ \mathcal{H}_{DW} =  \pi_{\phi}^i
\partial_{s_i}\phi + \pi_{\phi^\dagger}^i
\partial_{s_i}\phi^\dagger - L = 2 \pi_{\phi^\dagger}^1 \pi_\phi^2 + i
\pi_\phi^1 \square_1 \phi + i \pi_\phi^2 \square_2 \phi + h.c. .$$
One then proceeds to re-write it as a sum of two contributions,
each uniquely associated to a single time parameter:
$\mathcal{H}_{DW} = \mathcal{H}_1 + \mathcal{H}_2$ , with
$\mathcal{H}_i = \pi_{\phi^\dagger}^1 \pi_\phi^2 + i \pi_\phi^i
\square_i \phi + h.c$.

The Hamiltonians governing the \lq time evolution\rq\ with respect
to the different time directions identified by each variable $s_i$
are then defined by integration over independent hypersurfaces,
each orthogonal to a different time direction, e.g. $H_1 = \int
ds_2 dg_i \mathcal{H}_1$. Each $H_i$ results in being independent
of time $s_i$, as one would expect.

One can then proceed, after suitable decomposition in modes of
fields and momenta, the definition of (a GFT-adapted version of)
the covariant Poisson brackets, etc, to the canonical quantization
of the theory, with the definition of a Fock structure on the
space of states. In this way, one is able to make precise the
intuition of GFTs being a dynamical theory of creation and
annihilation of fundamental quanta of space. We refer once more to
\cite{ioJimmy} for the results of this analysis, among which we
mention only the interesting interplay between statistics and
group structure in GFTs, that causes fields on a Riemannian group
$G$ to be quantized as bosons, and fields on a Lorentzian group
$G$ to be quantized, necessarily, as fermions, in order to
preserve positivity of the Hamiltonians $H_i$. However, this
result may depend rather crucially on the specific choice of
kinetic term, and so be model-dependent, but we believe that this
shows how the issue of field statistics in GFTs deserves to be
further analyzed in more precise and general terms.

This is only one of the many features of the polysymplectic
formalism for field theories, in general, and of its GFT
incarnation, in particular, that need to be developed and
clarified at both the mathematical and physical level.

They include, at the classical level, and for the general
polysymplectic formalism: the formulation of a covariant
Hamilton-Jacobi theory corresponding to it, and the study of its
implications for quantum field theory; an analysis of the
relationship between the DeDonder-Weyl polysymplectic field theory
and other known covariant extension of ordinary Hamiltonian field
theory (see again \cite{kanatchikov} for a discussion); a complete
analysis of the exact reduction of formalism and results of
polymomentum Hamiltonian field theory to the usual single-time
Hamiltonian formalism; more technically, there many unsolved
issues concerning the algebraic structure induced on the space of
(horizontal) differential forms constituting the generalized phase
space  by Kanatchikov's definition of the (graded) Poisson bracket
for them (that seems to be a generalized type of Gerstenhaber
algebra); also, the general theory of conservation laws in
polymomentum mechanics, with the corresponding definition of
conserved currents and charges, the generalization of Noether
theorem, etc, is to be developed in more detail and its physical
consequences for know theories (and here we refer to ordinary
field theories as well as to GFTs) have to be analyzed, also given
the great physical importance attached to them.

At the quantum level, the state of the art is even more full of
open issues, given that work on this has really just started: for
the quantization of field theories based on the full
DeDonder-Weyl-Kanatchikov Hamiltonian theory we refer to
\cite{kanatchikov}, and subsequent work by the same author, for
the first steps; for the GFT-adapted formalism, the same
consideration apply, so we refer to \cite{ioJimmy}. We only
mention three open problems specifically related to the GFT
approach. The first is the general issue of conserved quantities
and symmetries for GFTs; this is crucial for a better
understanding of the appropriate scalar product in the space of
fields, or equivalently for the GFT definition of the kinematical
inner product for canonical 1st quantized quantum gravity wave
functions from the perspective of GFTs, and at the same time would
provide the basis for the analysis of symmetries at the quantum
level, and thus the consequent gauge fixing of Feynman amplitudes,
i.e. spin foam models; we will discuss this issue more in the
following. The second is related to the first: what is the GFT
analogue of the notion of anti-particles? Being strictly
intertwined with the complex structure of the field, one would
expect it to be linked with the orientation of the simplices
corresponding to the fundamental quanta of the GFT, and this would
match some insights coming from recent developments in spin foam
models; however, the whole issue is far from clear, and probably
is best addressed within an Hamiltonian context, like the one
outlined above. The third is the canonical derivation of GFT
quantum propagators, i.e. 2-point functions; in particular, one
would like to put on more solid grounds the choice of propagator
made in the construction of the spin foam models/Feynman
amplitudes corresponding to the models \ref{EF} and \ref{EFnr};
clearly, in light of the above discussion, this means providing,
among other things, a suitable definition, in an Hamiltonian
setting, of a \lq\lq multi-time-ordering\rq\rq of field operators.

\subsection{Symmetries}
We have mentioned above the open issue of symmetries in group
field theories. We now expand on it, trying to clarify what is
known and what is not known.

All the past work on symmetries in group field theories has
proceeded in a rather awkward way, from a field theoretic
perspective. Consider the model \ref{PR}, which is the only one of
which we understand reasonably well the symmetries and their gauge
fixing at the level of Feynman amplitudes. As said, the same
Feynman amplitudes can be derived from a discretization of the
continuum action and path integral of topological BF theory in 3
dimensions. One can \cite{PR1} identify the discrete analogue of
the continuum symmetries of BF theory: the translational and local
Lorentz symmetry, at the level of each GFT Feynman
diagram/simplicial complex, then devise the appropriate gauge
fixing procedure of the spin foam model/GFT Feynman amplitude, to
get rid of the redundant gauge degrees of freedom, and the
corresponding Faddeev-Popov determinant, to obtain in the end
fully gauge-fixed and finite GFT Feynman amplitude.

Now, first of all, this has been done, to date, -only- for this
specific GFT model, and not much is known about the relevant
symmetries of other models. Second, even in this case, from a GFT
perspective there is still much to be understood. In particular,
while the Lorentz symmetry has a clear GFT origin in the
invariance of the field under the diagonal action of the group G
on its 3 arguments, the GFT origin of the translation symmetry
remains mysterious\footnote{And a better understanding of this
translation symmetry in GFTs would be of great relevance from a
quantum gravity point of view since to has been shown \cite{PR1}
that it is strictly related to the discrete Bianchi identities on
the simplicial complex dual to the GFT Feynman diagram.}.

On top of this, as said, it is rather obvious that this is a
cumbersome way of proceeding, from a purely field theory
perspective: one would study directly the symmetries of the GFT
action that originates the Feynman amplitudes under consideration,
and derive the relevant perturbative identities between Feynman
amplitudes and n-point function, rather than try to guess what
these symmetries are from an analysis of the individual Feynman
amplitudes, which are effectively path integrals for the
corresponding single and multi-particle theories.

This is indeed what needs to be done and understood in all
mathematical and physical details. As mentioned, the natural
starting point would be the Hamiltonian analysis of symmetries fr
the various GFTs, and the derivation of the corresponding
conserved currents and charges. But one can as well remain at the
Lagrangian level and be concerned only with the perturbative
expansion of GFTs in terms of spin foam models; in this case, one
should identify the various symmetry transformations of the fields
in the GFT action, for any specific model considered, and derive
from them the corresponding Ward identities for the n-point
functions. These would be exactly the identities between spin foam
amplitudes that can (and were) in some cases discovered by direct
analysis of the classical discrete theory from which the same
amplitudes, can be derived. Both the general formalism for
deriving Ward identities and explicit examples of symmetry
analysis for specific GFT models are yet to be developed.

Notice that this is far from a trivial task, even at the classical
level, as one needs to work out a suitable generalization of
Noether theorem, adapted to GFTs, that overcomes the difficulties
posed by presence of several time variables (leading necessarily
to a polymomentum formalism), and by the product form of the
general kinetic operator, as well as by the higher-order in the
\lq\lq spacetime\rq\rq derivatives that results from it, even for
simple choices of the kinetic term.

One more reason why this would be of interest is that, as it was
stressed forcefully by P. Cvitanovic in his lectures at the
workshop, the presence of symmetries and of the consequent Ward
identities in a field theory affects greatly, among other things,
the growth rate and convergence properties of the corresponding
perturbative series, and thus in the GFT case it can tell us a lot
about the mathematical structure and properties of the sum over
topologies, as well as over geometries, implicit in their
expansion in Feynman diagrams.

\subsection{Combinatorial structure of the Feynman diagrams}
Let us now move on, indeed, to the discussion of the open problems
concerning the Feynman diagrams themselves. These have to do with
their combinatorial structure mainly, and can be motivated by the
need, and at the same time the chance that group field theories
offer, to re-phrase quantum gravity questions in purely (quantum)
field theoretic terms, and to tackle them with (quantum) field
theoretic tools.

To start with there is a urgent need to clarify the general
combinatorial structure of GFT Feynman diagrams, by developing the
basic concepts of ordinary quantum field theory in this new and
rather peculiar context. What are straightforward questions in
QFT, become a bit less straightforward ones in GFTs. For example:
what are the 1-particle irreducible diagrams, now that lines of
propagation are actually formed by several strands (equivalently,
what is the generalization of 1PI to 2-complexes)? and what are
their properties?  what is the (approximate) form of the 1-loop
effective action for the various GFTs? what is the exact
combinatorial content of the Dyson-Schwinger equations in this
setting?

If these are still rather simple questions (although may involve
technical or formal complications), as soon as we try to unravel
in more explicit term the combinatorial structure of GFT Feynman
diagrams with respect to their dual picture as simplicial
complexes, things start to complicate considerably.

Even if every 2-complex arising as a GFT Feynman diagram can be
understood as topologically dual to a D-dimensional simplicial
complex, this would not be, in general, a simplicial {\it
manifold}. In fact, the data attached to GFT Feynman diagrams, nor
the feynmanological rules for their construction, do not constrain
the neighbourhoods of simplices of dimensions from (D-3) downwards
to be spheres. This implies that in the general case, the
resulting simplicial complex, obtained by gluing D-simplices along
their (D-1)-faces, would correspond to a {\it
  pseudo-manifold}, i.e. to a manifold with {\it conical singularities}
\cite{DP-P}. The issue does not arise in D=2, where all GFT
Feynman diagrams (combinatorially the same as those of simple
matrix models) are dual to simplicial manifolds, if the
orientation condition is satisfied. Neglecting for the moment the
issue of whether this is a problem from a physical
perspective\footnote{As pointed out by one of the participants
during the workshop, after all, we do not know whether or not
spacetime {\it really} possesses conical singularities in the
microscopic regime....}, or whether on the contrary it is possible
to give some physical meaning to these singularities, one remains
with the task of analyzing these configurations from a purely
mathematical point of view. A precise set of conditions under
which the GFT Feynman diagrams correspond to manifolds is
identified and discussed at length in \cite{DP-P}, both at the
level of simplicial complexes and of the corresponding dual
2-complexes, in D=2,3,4. All the relevant conditions can be
checked algorithmically on any given Feynman graph. However, we
feel it would be good to build upon the analysis of \cite{DP-P},
in two main directions: 1) try to identify a suitable
reformulation of the found conditions, or an alternative but
equivalent set of conditions, that would make the quantum field
theoretic interpretation and role of the pseudo-manifold
configurations more transparent; 2) if and once this can be done,
find how to impose these conditions at the GFT level (a sort of
superselection rules?) or construct suitably constrained GFT, that
would generate only manifold-like complexes in their Feynman
expansion. This last task makes sense, of course, only if the
first does not prove that, because of their field theoretic
interpretation, pseudo-manifold configurations are in fact needed
for consistency at the quantum level (in the same sense, for
example, as loop diagrams are in ordinary field theory). It may
also turn out that non-manifold-like configurations can not be
removed but are instead suppressed in certain sectors of the
theory, in specific models, as for example happens in some 3d
tensor models \cite{gross}.

Another wide landscape of interesting questions opens up when
considering the topological structure of the GFT Feynman diagrams.
As discussed above, the sum over GFT Feynman diagrams includes a
sum over all simplicial topologies as well as over all simplicial
decompositions of the same topology. In D=2 different topologies
are weighted by a single topological invariant, the Euler
characteristics, and one can then try to express the GFT Feynman
amplitudes as a function of this topological parameter, and then
identify the sector of the theory or of the parameter space in
which, say, the trivial topology dominates. In matrix models,
indeed, where the Feynman amplitudes have a simpler form (as
discussed above, they correspond to the truncation of the simple
GFT \ref{mm} to a fixed representation J), one can easily show
that non-trivial topologies are suppressed in the limit of
infinite matrix dimension (large J). No analogue of this result in
the full D=2 GFTs is known. In $D\geq 3$ topologies are not
classified so the situation is much more intricate. A first guess
at how non-trivial topologies enter the GFT Feynman expansion is
that they only appear beyond tree level, i.e. in the quantum
regime; in fact, in matrix models the expansion in the matrix
dimension can also be understood as a loop expansion and confirms
the above expectation. This was confirmed also for generic GFTs in
\cite{laurentgft} by analyzing the Dyson-Schwinger equations: at
tree level only diagrams of trivial topology appear, and the order
of loops can then be related to the number of handles in the
simplicial complex dual to the Feynman diagram. Moreover, by
suitable re-scaling of the field, the number of loops could be
related to (a power of) the GFT coupling constant, that thus
acquires a possible interpretation as the parameter governing
topology change; we refer to \cite{laurentgft} for more
details\footnote{Much remains to be understood concerning this
correspondence, anyway, as one can easily find high order (in
$\lambda$) diagrams which still correspond to trivial topology.}

The number of handles is, however, just one of the quantities that
can characterize simplicial topology, and a more in dept analysis
of the relation between the topological properties of the
simplicial complexes appearing in the Feynman expansion and their
field theoretic interpretation would be very much welcome. And
more generally, one would also like to be able to identify more
clearly the dependence of the GFT Feynman amplitudes (for generic
dimension, and for non-trivial models) on the topology of the
underlying diagram. This would help greatly in the attempt to gain
control over the sum over topologies implicit in the GFT
perturbative expansion.

For example, just as one can relate handles and quantum loops, is
there a QFT interpretation of other known topological invariants
or for other topological properties of simplicial complexes?

The model \ref{PR}, for example, generates Feynman amplitudes that
are topological invariants themselves \cite{PR1}, meaning that
they evaluate to the same number (after appropriate gauge fixing
and regularization) on any element of equivalence class of
simplicial complexes related by Pachner moves, while they still
provide a different amplitude for topologically inequivalent
simplicial complexes. Can one identify the field theoretic feature
of this model (and of its higher dimensional equivalents, whose
amplitudes share the same property) that can be seen as the origin
of this property of its amplitudes? If this can be done, can we
use this new insight to construct new topological invariants of
simplicial complexes by field theoretic means?

Even remaining at the level of trivial topology, e.g. say we just
consider diagrams with the cylindrical topology
$\Sigma\times\mathbb{R}$, there are further structures that one
would like to identify in the diagrams and characterize in
field-theoretic terms. In the modern version of the dynamical
triangulations approach \cite{CDT}, namely {\it causal dynamical
triangulations}, one finds that imposing additional combinatorial
restrictions to the simplicial complexes summed over in the
definition of the quantum gravity path integral, the continuum
properties of the same are drastically improved and one can
recover many of the wanted properties of a continuum spacetime
from its purely combinatorial quantum definition, e.g. its
spectral dimension \cite{CDT}. These additional restrictions
include the presence of a fixed foliation of the triangulations
and the absence of branching (baby universe) configurations with
respect to this foliation. Given that the GFT partition function
can be re-expressed, in perturbative expansion, as a sum over
triangulations, weighted by Feynman amplitudes, what is the field
theory interpretation of these combinatorial restriction? or, can
one identify specific GFT models or general properties of GFT
Feynman amplitudes that would lead to a strong suppression, if not
the absence, of configurations not satisfying these restrictions?

\subsection{Renormalization}
We now come to the important issue of renormalization of group
field theories, the importance of which we certainly do not need
to stress, also given that all known GFT models, at present, are
{\it defined} by their perturbative expansion.

Once more, this is almost completely unexplored territory. The
whole question of perturbative renormalization of the various GFT
models has not been analyzed in any detail, nor any general scheme
for performing the perturbative renormalization of GFTs has been
developed. Once more, it is the peculiar combinatorial structure
of the GFT Feynman diagrams, and the non-local pairing of
variables in the vertex term in the action, that makes the whole
issue of divergences and renormalization at the same time very
intricate and challenging. For any given Feynman diagram, and
after gauge fixing, the sum over geometric data has two potential
sources of divergences: depending on the kinetic term (propagator)
one chooses, one can have a potential divergence for each loop,
i.e. each 2-cell, but also one has a potential divergence for
every \lq bubble\rq of the GFT Feynman diagram, i.e. for every
closed surface of it identified by a collection of 2-cells glued
together along common edges. This is, in a sense, the true GFT
analogue of loop divergences of usual QFT. In addition to this \lq
double-layer\rq structure of potential divergences, there are
other combinatorial peculiarities of GFTs that make the
conventional wisdom (including, for example, simple counting of
the degree of divergence of a diagram) less easily applicable, and
a brand new scheme of analysis badly needed. One example is the
fact that a closer look at the way degrees of freedom are
propagated within a Feynman diagram reveals a sort of \lq
propagation-only\rq structure, in the sense that also within
vertices of interaction the field theory degrees of freedom are
simply re-routed, i.e. propagated, and there is no coincidence at
the same \lq\lq point\rq\rq of more than two field variables. This
may mean that a regularization of the GFT propagators may suffice
to cure loop divergences. However, also this needs a more careful
study. Certainly, due to this combinatorial peculiarities, usual
results on the non-renormalizability of $\phi^n$ theories for high
enough $n$ have to be at least reconsidered, to check that they
still apply in the GFT framework.

In fact, whether the GFT amplitudes are divergent or not depends
on the specific model, even for the same polynomial order in the
interaction term. For example, the model \ref{PR} turns out to be
finite after gauge fixing, while the most natural definition of
the group field theory for the Barrett-Crane spin foam model for
4d gravity \cite{SF}, for example, presents indeed bubble
divergences, and the Perez-Rovelli modification of it \cite{SF},
producing a different version of the same model based on a GFT of
the same order and same combinatorics, but different symmetries,
possesses again {\it finite} Feynman amplitudes, i.e. it is {\it
perturbatively finite} without the need for any regularization,
even before any gauge fixing.

Once more, one has to develop thus a general theory of
perturbative renormalization for GFTs, that would allow to unravel
first the combinatorial structure of GFT divergences, and then to
regularize them away. On the one hand, conventional regularization
and renormalization techniques would be the first thing to try as
they look like the most efficient way of performing explicit
computations and extract physical results from a field theory;
indeed, the development and use of their GFT-adapted analogues
would be of great value. On the other hand, the Hopf algebra
approach to renormalization \cite{renormhopf} have proven to be
especially suitable for capturing and elucidating the
combinatorial structure of Feynman diagrams and of their
divergences; therefore it is a natural guess and hope that one can
apply similar techniques to GFTs, where usual tools may be less
powerful exactly because of the combinatorial intricacies of the
GFT Feynman diagrams (some of which have been highlighted in the
previous section).

That the Hopf algebra techniques developed by Kreimer and Connes,
among many others, can be the correct language to tackle the issue
of GFT renormalization is suggested also by recent work on the
renormalization of spin foam models, in particular by the work of
\cite{markopoulou}. Here the conceptual setup and the perspective
on spin foam models were rather different from the one presented
in the present article, and group field theories were really not
part of the picture. Instead, spin foam models were studied as
background independent discrete quantum gravity path integrals (or
statistical mechanical state sum models), and the task was to
develop a coarse graining and renormalization procedure, a
background independent analogue of the usual one used in lattice
statistical mechanics and statistical field theory, that would
bypass the difficulties coming from the absence of a fixed lattice
geometry (e.g. fixed lattice spacing) and could be used for
lattices of arbitrary combinatorial structure, as spin foams. The
most natural language was found to be, in fact, that of Hopf
algebra renormalization and a {\it Hopf algebra of spin foams}
(more precisely, a Hopf algebra of {\it partitioned} spin foams)
was defined, together with a Hopf algebra of (parenthesized) spin
foam weights (amplitudes), based on the identification of
appropriate {\it subfoams}. As in the Connes-Kreimer approach, it
is the antipode of the two algebras that plays a crucial role in
the definition of renormalization group transformations. Indeed,
Markopoulou went on defining exact and approximate block
transformations on spin foams, based on the discovered Hopf
algebra structure, and in particular on its antipode, and
suggested that this could be the correct starting point for the
analysis of renormalization of known spin foam models.

This approach acquires a new light  and, in a sense, a further
justification from a group field theory perspective\footnote{This
was indeed noted already in \cite{markopoulou}}. Recall that spin
foams are nothing more than the Feynman diagrams of group field
theories, and spin foam models are their Feynman amplitudes. The
definition of a renormalization procedure for spin foam models
would therefore amount to a definite prescription for perturbative
renormalization of group field theories. From this standpoint,
then, the fact that the Kreimer-Connes Hopf algebra approach to
renormalization, originally developed for perturbative
renormalization of QFTs, can be adapted to spin foam models seems
only natural. The tasks however are: to identify the
Kreimer-Connes algebra of Feynman diagrams for GFTs and compare it
with the algebra of spin foams as defined by Markopoulou (as a
first step, this involves comparing GFT 1PIs to the subfoam
structure proposed in \cite{markopoulou}); go on identifying the
corresponding Hopf algebra of GFT perturbative renormalization,
and again compare it with the one of \cite{markopoulou}; clarify
the role of GFT gauge symmetries within it; apply it to specific
GFT models. Given the large number of interesting mathematical
results coming out of recent work on QFT Hopf algebra
renormalization, we expect further work along these lines in the
GFT framework to be particularly exciting and rewarding, also
considering on the one hand the dual simplicial spacetime
interpretation of GFT diagrams, and on the other hand the
relevance of any result so obtained for quantum
gravity\footnote{In fact such an analysis would represent one
direct way of tackling the issue of the continuum limit of spin
foam models and of their relation with continuum Einstein's
gravity.}.

In the context of renormalization, let us also mention that it
would be of great interest to go beyond the perturbative level and
develop an exact Wilsonian renormalization group analysis of group
field theories. On top of providing important information on the
renormalizability of GFTs, it would represent an even more (with
respect to the perturbative renormalization of their Feynman
diagrams/spin foam models) powerful tool for the study of their
continuum limit and relation with classical General Relativity.

Another related issue, concerning the perturbative expansion of
GFTs in Feynman diagrams/spin foam models, is that of summability
of the perturbative series. Of course, even though, from the
interpretation of it as a discretized version of a quantum gravity
path integral, one may hope it to be finite, this is not such a
reasonable expectation from a purely field theory perspective,
which is the one we would like to advocate here. However,
summmability is a more limited but perfectly reasonable hope for a
QFT perturbative series. Notice that such a property would still
be rather remarkable, even though only if achieved in a physically
meaningful way, from a quantum gravity perspective, given that it
amounts to the possibility of gaining (non-perturbative) control
over a sum over {\it all simplicial geometries and all simplicial
topologies}. Also this issue has been largely unexplored, both in
general terms and for specific GFT models. The only (important)
result in this direction obtained so far concerned the model
\ref{PR} and is the following\cite{laurentborel}: a simple
modification of the GFT action \ref{PR} gives a model whose
perturbative expansion is Borel summable. The modification amounts
to adding another vertex term of the same order to the original
one, to give:

\begin{eqnarray*}
+\frac{\lambda}{4!}\prod_{i=1}^{6}\int_{SU(2)}dg_i\left[
\phi(g_1,g_2,g_3)\phi(g_3,g_4,g_5)\phi(g_5,g_2,g_6)\phi(g_6,g_4,g_1)+
\,\delta\,
\phi(g_1,g_2,g_3)\phi(g_3,g_4,g_5)\phi(g_4,g_2,g_6)\phi(g_6,g_5,g_1)\right],
\end{eqnarray*}
with $\mid \delta\mid < 1$.

The new term corresponds simply to a slightly different recoupling
of the group/representation variables at each vertex of
interaction, and geometrically to the only other possible way of
gluing 4 triangles to form a closed surface, i.e. to form a \lq\lq
pillow\rq\rq instead of a regular tetrahedron. In turn this \lq\lq
pillow\rq\rq configurations are equivalent to two tetrahedra glued
together along -two- common triangles, instead of one\footnote{It
would be interesting to understand in more detail the field
theoretic interpretation of these configurations.}. Even if the
above modification of the Boulatov GFT model has no clear physical
interpretation yet from the quantum gravity point of view, e.g. in
terms of gravity coupled to some sort of matter, it is indeed a
very mild modification, and most importantly one that one would
expect to be forced upon us by renormalization group-type of
argument, that usually require us to include in the action of our
field theory all possible terms that are compatible with the
symmetries. What the other terms would be and what their effect on
the perturbative series is another interesting open issue,
together with the (more physically important) possibility of
obtaining a similar result for any of the known GFT models of
quantum gravity in $D=4$.

\section{Conclusions}

In this article, we have introduced the group field theory
formalism, trying to clarify its relevance for quantum gravity
research, as well as key aspects and peculiarities of GFT models,
defining combinatorially non-local field theories on the one hand,
and a generalization of matrix models and field theories of random
surfaces on the other. Most importantly, we have presented and
discussed many outstanding open issues and unanswered questions
that we thought could be of special interest to mathematical
physicists and mathematicians working on quantum field theories in
general, and in particular those interested in the related
perturbative combinatorics and renormalization\footnote{The
selection, of course, has been operated from the particular (and
certainly limited) perspective of a theoretical physicist working
on quantum gravity, and this has affected also the style of the
discussion. Being unavoidable, we do not feel the need to
apologize for this.}. Our expectation is that GFTs can represent
for mathematicians and mathematical physicists a very valuable
playground and toolbox, and an important source of insights and,
obviously, amusement, also in light of the immense corpus of
mathematical results, tools and applications that has already
developed from work on matrix models and random surfaces (and
which is by all means still growing). Adding to this the mentioned
relevance of all of these open issues for the construction of a
satisfactory theory of quantum gravity, the urge to join the
current efforts of many theoretical physicists working in this
area should be, we hope, unrestrained.....

\section*{Acknowledgements}
We would like to thank the organizers of the Conference on
Combinatorics and Physics for their kind invitation and for having
made it a very enjoyable and profitable event.

\end{document}